\documentclass[showkeys]{revtex4}

\usepackage{dcolumn}
\usepackage{amssymb}
\usepackage{graphicx}

\begin{document}


\newcommand \be {\begin{equation}}
\newcommand \ee {\end{equation}}
\newcommand \bea {\begin{eqnarray}}
\newcommand \eea {\end{eqnarray}}
\newcommand \ve {\varepsilon}
\newcommand \la {\langle}
\newcommand \ra {\rangle}

\title{How far can stochastic and deterministic views be reconciled?}

\author{Eric Bertin}

\affiliation{Universit\'e de Lyon, Laboratoire de Physique, Ecole Normale Sup\'erieure de
Lyon, CNRS, 46 All\'ee d'Italie, F-69007 Lyon, France.}

\begin{abstract}
In this short note, we try to provide the reader with a brief pedagogical
account of some similarities and differences between stochastic
and deterministic processes. A short presentation
of some basic notions related to the mathematical description of stochastic
processes is also given.
Our main aim is to illustrate the somehow surprising fact that the gap
between the behaviour of stochastic and deterministic processes might,
from a practical perspective, be much smaller than a priori expected.
\end{abstract}


\maketitle

\section{Introduction}

Stochasticity and randomness appeared in physics nearly one century ago,
with the emergence of statistical physics and quantum mechanics
\cite{Griffiths,Rae,VanKampen}
--see also the paper by M.~Le Bellac in this special issue.
Randomness was at this stage essentially confined to the atomic scale,
and a predictable behaviour was still assumed to hold at the macroscopic level.
In the last fifty years, it was however recognized that even macroscopic
and deterministic systems, with a small number of degrees of freedom,
could behave in an unpredictable and apparently random manner;
such systems have been called chaotic \cite{Berge}.
Stochasticity also appears nowadays as an essential ingredient in the
dynamics of biological systems \cite{Kupiec,CHC}, as witnessed by the
present Special Issue on stochastic processes in cell biology.

In this note, we wish to illustrate, using a few simple mathematical examples,
that chaotic and stochastic dynamics may look quite similar in certain cases,
in spite of important conceptual differences.
In addition, deterministic chaotic systems can in some situations be modelled
by stochastic processes, as we shall see in the last section.
This suggests that the gap between stochastic and deterministic (chaotic)
dynamics might in some cases, from a practical perspective,
be smaller than a priori expected.

The article is organised as follows. In section~\ref{sec-stoch},
we present some basic notions on the mathematical formalism
used to describe stochastic processes.
In section~\ref{sec-deter}, we illustrate on a simple case the emergence
of chaos in a deterministic system when varying a control parameter,
as well as the notion of ``chaotic walk'', the deterministic analog
to random walks.
Finally, section~\ref{sec-compar} discusses the possibility to
use stochastic models in order to mimick the coarse-grained dynamics
of a deterministic process.

\section{Basic notions on stochastic processes}
\label{sec-stoch}

In this first section, we tentatively provide the reader with
a brief and pedagogical presentation of the basic mathematical
formalism describing stochastic processes, a knowledge useful
to better grasp the essence of stochasticity.
The interested reader is referred to standard textbooks, like
Ref.~\cite{VanKampen}, for more details.

\subsection{Master equations}

Let us start by considering the case of discrete time stochastic processes,
that is stochastic processes with events randomly occuring
between integer times $t=0,1,2,...$
The different configurations accessible to the process are labeled
by an integer $n$.
The process is described by transition probabilities $T(n' \to n)$
for all $n,n'$, which gives the probability that the configuration is $n$
at time $t+1$ given that it was $n'$ at time $t$.
Note that the configuration does not necessarily change at each step,
the probability $T(n' \to n')$ is generally non zero.

The statistics of the process is described the probability $P_n(t)$ to be in state $n$ at time $t$. This probability evolves according to the following
discrete-time master equation
\be \label{ME-discrete-time}
P_n(t+1) = \sum_{n'} T(n' \to n) P_{n'}(t)
\ee
which simply accounts for the balance of probabilities: the probability
to be in state $n$ at time $t+1$ results from all the transitions from any
configurations $n'$ to configuration $n$ between times $t$ and $t+1$,
weighted by the transition probability $T(n' \to n)$ and by the probability
$P_{n'}(t)$ to occupy configuration $n'$ at time $t$.
It is important to note that we focus here on processes where no
memory of previous configurations at time $t-1$, $t-2$,..., is present.
Such processes are called ``Markovian stochastic processes''.

The continuous time case, which is the most commonly used in practical
modelling, can be obtained from the discrete time case by taking the
limit of infinitely short time steps. We thus choose very short time steps
$\Delta t = dt$, instead of $\Delta t = 1$ as previously.
It is then necessary to specify the behaviour of the transition probability
with the time step $dt$. One expects in particular that the shorter
the time interval $dt$, the smaller the transition probability to a different
configuration in this interval. In other words, in a short time interval,
the process ``has not enough time'' to change configuration, and most likely
remains in the same one.
A natural choice for the transition probabilities is thus
\bea
T(n' \to n) &=& W(n' \to n)\, dt \qquad \mathrm{if} \quad n' \ne n\\
T(n \to n) &=&  1 - \sum_{n' \ne n} W(n' \to n)\, dt + \mathcal{O}(dt^2)
\eea
This choice indeed ensures that the different probabilities sum up to one,
as expected. The new quantity $W(n' \to n)$ that has been introduced
is called a ``transition rate'' (i.e., a transition probability per unit
time). This quantity plays a key role in the description of continuous time
stochastic processes.

The evolution of the probability $P_n(t)$ to be in configuration $n$
at time $t$ can be deduced from Eq.~(\ref{ME-discrete-time}).
Expanding $P_n(t+dt)$ to first order in $dt$ yields
\be
P_n(t+dt) = P_n(t) + \frac{dP_n}{dt}\, dt + \mathcal{O}(dt^2).
\ee
Gathering terms linear in $dt$, one finds the
continuous time master equation
\be \label{eq-ME}
\frac{dP_n}{dt} = \sum_{n'} \Big[ -W(n\to n') P_n(t)
+ W(n'\to n) P_{n'}(t) \Big].
\ee

\begin{figure}
\centering\includegraphics[width=0.35\textwidth,clip]{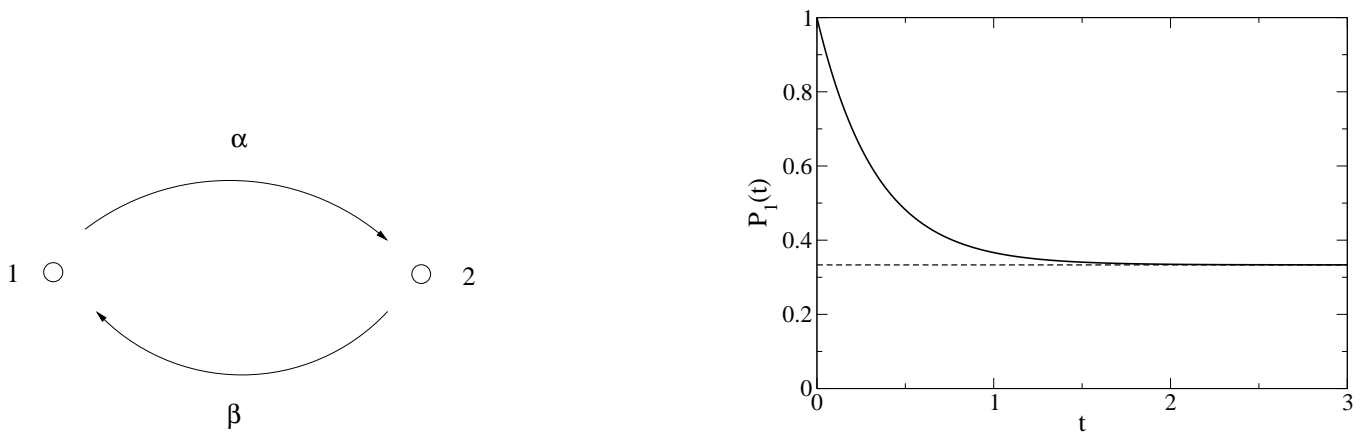}
\hfill
\centering\includegraphics[width=0.45\textwidth,clip]{2state.eps}
\caption{Left: Illustration of the two-state model.
Right: Relaxation of the probability $P_1(t)$ (full line) to its asymptotic
equilibrium value (dashed line) in the two-state model
(parameters: $\alpha=2$, $\beta=1$).}
\label{fig-2state}
\end{figure}

As a simple illustration, let us consider a process with only two
configurations, $n=1$ and $2$.
The transitions rates are $W(1\to 2)=\alpha$ and $W(2\to 1)=\beta$,
see the left panel of Fig.~\ref{fig-2state}.
The master equation corresponds to the following set of equations
\bea
\frac{dP_1}{dt} &=& - \alpha P_1 + \beta P_2 \\
\frac{dP_2}{dt} &=& - \beta P_2 + \alpha P_1
\eea
Taking into account the relation $P_1+P_2=1$, these two equations are found
to be equivalent, and can be reformulated as a single equation, namely
\be
\frac{dP_1}{dt} = - (\alpha+\beta) P_1 + \beta.
\ee
The solution of this equation reads
\be
P_1(t) = \frac{\beta}{\alpha+\beta} +
\left(P_1(0)-\frac{\beta}{\alpha+\beta}\right) e^{-(\alpha+\beta)t}
\ee
where $P_1(0)$ is the initial value of the probability at time $t=0$.
One observes that $P_1(t)$ relaxes to the stationary value
$P_{\mathrm{st}}$ as time elapses.
An illustration of the behaviour of $P_1(t)$ is provided in the right panel
of Fig.~\ref{fig-2state}.

Coming back to the general (continuous time) master equation
given in Eq.~(\ref{eq-ME}),
it is interesting to try to characterize the stationary solutions,
that is the solutions that are independent of time,
corresponding to $dP_n/dt=0$.
It follows that for all $n$,
\be
\sum_{n'} \left[ -W(n\to n') P_n + W(n'\to n) P_{n'} \right] = 0.
\ee
A specific case of interest is when all terms in the sum are equal to zero,
namely for all $n$ and $n'$
\be
-W(n\to n') P_n + W(n'\to n) P_{n'} = 0
\ee
This situation is called detailed balance, and it turns out to be
useful in practice to build models with desired stationary probabilities
\cite{VanKampen}.

\subsection{Example of the random walk on a line}

The one-dimensional random walk, describing for instance the stochastic
displacement of a particle on a line, can be defined as follows:
\be \label{eq-rw}
x_{t+\tau} = x_t + \epsilon_t, \qquad \epsilon_t = \pm a,
\ee
where $\tau$ is the constant time step.
The values $+a$ and $-a$ are chosen at random with equal probabilities.
In order to perform averages, one needs to use the so-called ensemble
averages, corresponding to averages over many different realizations of the
random walk. The average values obtained in this way are thus time-dependent,
and should not be confused with time-averages.

It is easy to check that the average value of the position of the
random walk is zero at all time, $\langle x_t \rangle =0$,
if the random walk initially starts from the position $0$.

A more interesting quantity to characterize the statistical properties
of the walk is the mean square displacement $\langle x_t^2 \rangle$.
This quantity can be computed as follows.
Taking the square of Eq.~(\ref{eq-rw}), one finds
\bea
x_{t+\tau}^2 &=& (x_t + \epsilon_t)^2\\
&=& x_t^2 + 2x_t \epsilon_t + \epsilon_t^2.
\eea
Averaging over many realizations of the process yields
\be
\la x_{t+1}^2 \ra = \la x_t^2\ra + 2 \la x_t \epsilon_t\ra +
\la \epsilon_t^2 \ra.
\ee
We first note that $\epsilon_t = \pm a$ implies $\epsilon_t^2=a^2$,
so that $\la \epsilon_t^2 \ra = a^2$.
Then, given that $\epsilon_t$ is statistically independent of $x_t$,
one has $\la x_t \epsilon_t\ra = \la x_t \ra \; \la \epsilon_t\ra = 0$.
As a result,
\be
\la x_{t+\tau}^2 \ra = \la x_t^2\ra + a^2
\ee
so that, assuming $x_0=0$, one eventually obtains
\be
\la x_t^2\ra = \frac{a^2}{\tau}\, t,
\ee
which means that the mean square displacement is linear in time.
This property is an important feature of random walk, called diffusive
behaviour. To emphasize the non-trivial character of this property,
let us indicate that the typical distance travelled by the walk after
time $t$ is given by $\sqrt{\la x_t^2\ra}$, and is thus proportional
to $\sqrt{t}$, which at large time is much smaller than the distance $v_0 t$
travelled by a particle moving at constant velocity $v_0$.

In the continuous time case, the random walk is characterized by
transition rates $W(n \to n-1)=W(n \to n+1)=1/2\tau$,
and $W(n \to n')=0$ if $n'$ is different from $n \pm 1$.
A detailed statistical description of the random walk
is then obtained from the corresponding master equation, which reads
\be
\frac{dP_n}{dt} = \frac{1}{2\tau} (P_{n+1}+P_{n-1}-2P_n)
\ee
where $P_n(t)$ is the probability to be at position $n$ at time $t$.
At large times, the probability profile appears essentially continuous,
and can be approximated by a continuous space equation. Introducing
the variable $x=na$ (with $a$ the lattice spacing) and a function
$p(x,t)$ of the real variable $x$ satisfying $P_n(t)=a\, p(x,t)$, one obtains
the following continuous description, valid for times much larger
than the microscopic time constant $\tau$:
\be \label{eq-diff}
\frac{\partial p}{\partial t} = D\,\frac{\partial^2 p}{\partial x^2}.
\ee
This equation is called the diffusion equation, and also appears
in other fields of physics, like the diffusion of heat in a material.
The quantity $D=a^2/2\tau$ is called the diffusion coefficient.

\begin{figure}
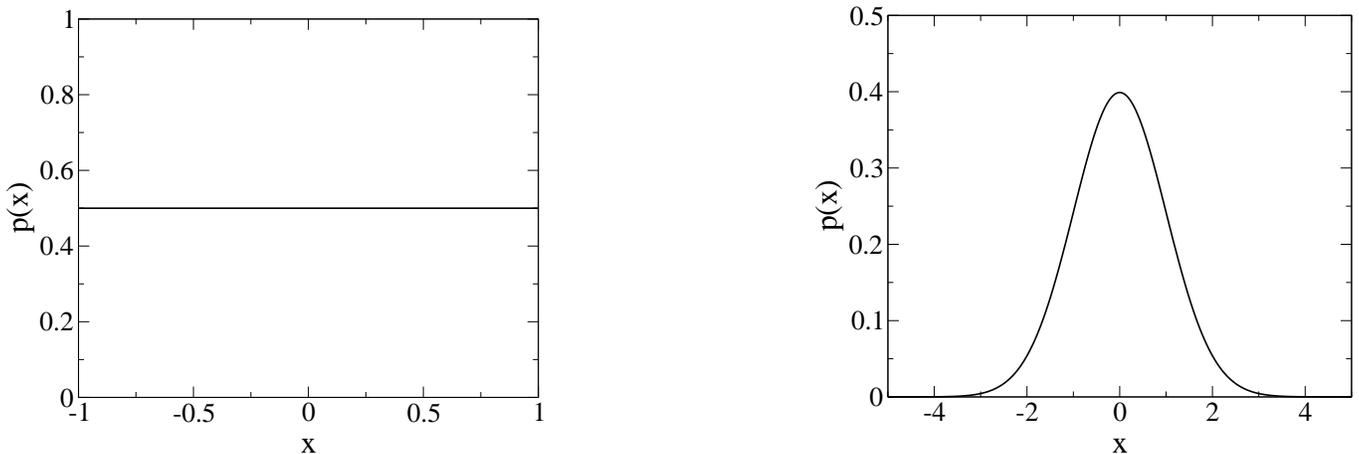

\includegraphics[width=0.4\textwidth,clip]{diff-unif.eps}
\hfill
\includegraphics[width=0.4\textwidth,clip]{diff-gauss.eps}
\caption{Illustration of the large-time behaviour of the probability
profile of the random walk. Left: diffusion on the finite segment
$[-1,1]$.
Right: diffusion on an infinite line, in the absence of confining
boundaries.}
\label{fig-diff}
\end{figure}

At very long time, the probability profile depends on the boundary
conditions.
If the walk is confined on a finite segment, the probability $p(x,t)$
tends to a spatially uniform distribution on this segment when
the time $t$ goes to infinity (left panel of Fig.~\ref{fig-diff}).
In contrast, if the walk diffuses on an unbounded domain,
it converges to a Gaussian shape that keeps broadening as time elapses,
as seen on the right panel of Fig.~\ref{fig-diff}.

Interestingly, this Gaussian shape of the distribution,
found by solving Eq.~(\ref{eq-diff}), can also be given an alternative
interpretation.
Starting from the discrete time formulation of the random walk
given in Eq.~(\ref{eq-rw}), one can reexpress $x_t$ as a sum of random
variables according to
\be
x_t = \sum_{t'=0}^{t-1} \epsilon_{t'} \,.
\ee
One can then apply the Central Limit Theorem, which states that
the sum of a large number of independent and identically distributed
random variables are distributed according to a Gaussian distribution
with a variance proportional to the number of terms.
One thus precisely recovers, without any extra calculation,
the result obtained above by solving the diffusion equation
Eq.~(\ref{eq-diff}).

Before concluding this brief introduction to the stochastic formalism,
it is worth mentioning a slight generalization of the diffusion equation,
called the Fokker-Planck equation. This equation describes the case
when the random walk is not necessarily symmetric, that is the
probabilities to go to the left and to the right are not necessarily equal.
This situation can be formulated mathematically as follows.
One considers a discrete time random walk defined by the recursion
relation $x_{t+1} = x_t + \epsilon_t$, where $\epsilon_t=+a$
with probability $\frac{1}{2}(1+aq_n)$, and $\epsilon_t=-a$
with probability $\frac{1}{2}(1-aq_n)$.
Denoting $q_n=Q(na)$ and using arguments similar to the ones considered
above to derive the diffusion
equation (\ref{eq-diff}), one obtains the following Fokker-Planck equation:
\be
\frac{\partial p}{\partial t}(x,t) = 
- \frac{\partial}{\partial x} \Big( 2D\, Q(x)\, p(x,t) \Big)+
D\, \frac{\partial^2 p}{\partial x^2}(x,t).
\ee
This equation turns out to be useful in many different modelling contexts,
from physics to chemistry and biology \cite{VanKampen}.

\section{Deterministic dynamics: from regularity to chaos}
\label{sec-deter}

\subsection{A simple deterministic process}
\label{sec-map-fx}

We now turn to the study of deterministic processes, with the aim
to compare them to stochastic processes under some circumstances.
A time signal is considered as deterministic if 
the knowledge of the value $x(t)$ of the variable considered
determines the value $x(t')$ of the variable at any later time $t'>t$.
A standard mathematical formulation for continuous time deterministic
processes is the following type of differential equation
\be \label{eq-ODE}
\frac{dx}{dt} = F\big(x(t)\big).
\ee
In the case of discrete time processes, one rather has:
\be
x_{t+1} = f(x_t).
\ee
Note that the discrete time case can be seen as a periodic sampling
of a continuous time process. In this case, the function $f(x_t)$
can be obtained by integrating the differential equation (\ref{eq-ODE})
between times $t$ and $t+1$.

\begin{figure}[t]
\centering\includegraphics[width=0.5\textwidth,clip]{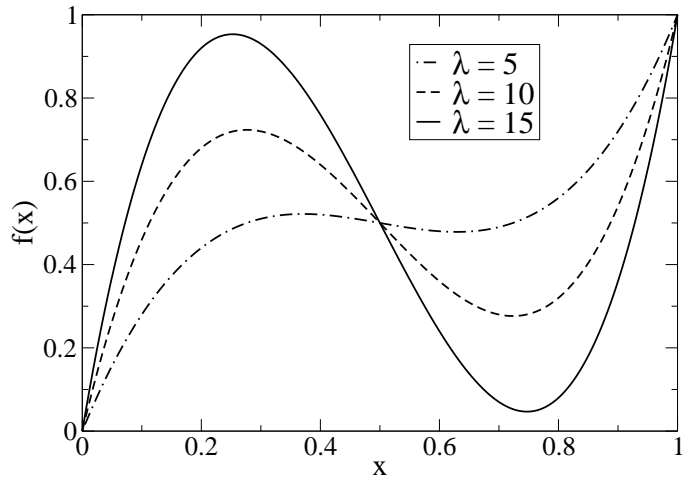}
\caption{Illustration of the shape of the function $f(x)$
for $\lambda=5$ (dot-dash), $\lambda=10$ (dashed line) and $\lambda=15$
(full line).}
\label{fig-map-fx}
\end{figure}

A simple and illustrative example is provided by the discrete
time deterministic process $x_{t+1} = f(x_t)$, with the function $f(x_t)$
given by
\be \label{eq-map-fx}
f(x) = x\left[\lambda \left(x-\frac{1}{2}\right)(x-1)+1\right].
\ee
The function $f(x)$ depends on a parameter $\lambda$,
and its shape is illustrated
in Fig.~\ref{fig-map-fx} for three different values of $\lambda$.

\begin{figure}[t]
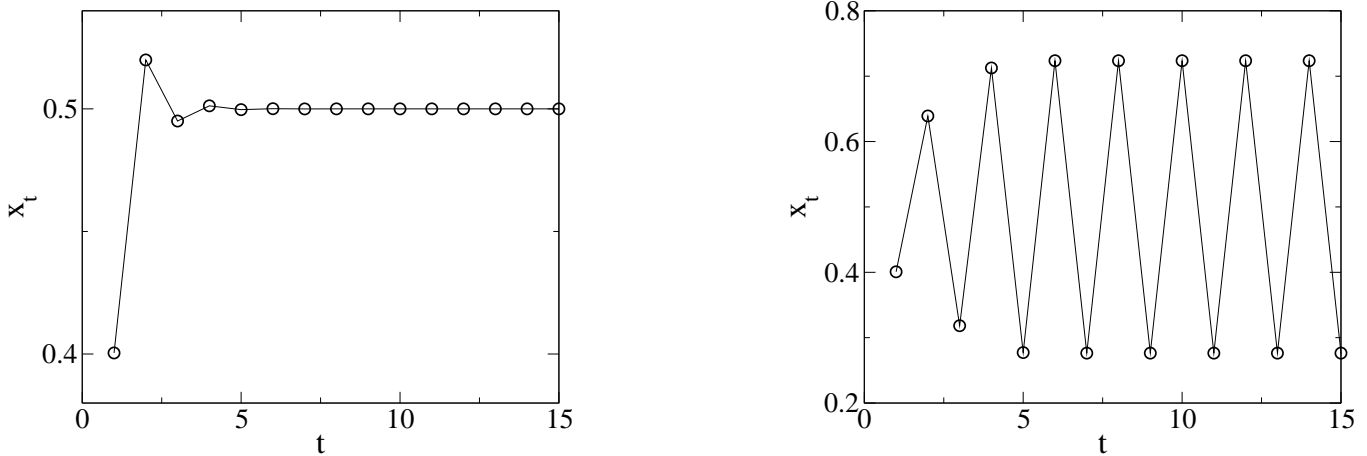

\includegraphics[width=0.42\textwidth,clip]{incr5.eps}
\hfill
\includegraphics[width=0.42\textwidth,clip]{incr10.eps}
\caption{Left: Convergence to a fixed point for $\lambda=5$.
Right: Convergence to a limit cycle (oscillation between two different points)
for $\lambda=10$.}
\label{fig-fx5-10}
\end{figure}

\begin{figure}[t]
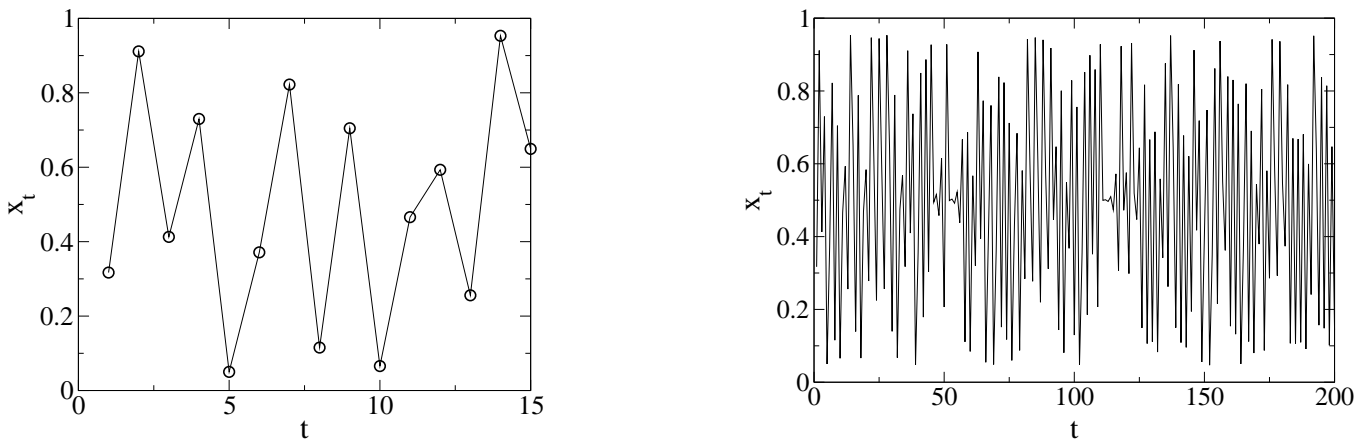

\includegraphics[height=0.32\textwidth,clip]{incr15.eps}
\hfill
\includegraphics[height=0.32\textwidth,clip]{incr15-lg.eps}
\caption{Case $\lambda=15$, showing a chaotic behaviour.
Left: Same time window as on Fig.~\ref{fig-fx5-10}.
Right: Larger time window.}
\label{fig-fx15}
\end{figure}

The behaviour of the deterministic process $x_{t+1} = f(x_t)$
is illustrated on Figs.~\ref{fig-fx5-10} and \ref{fig-fx15}
for different values of $\lambda$. For $\lambda=5$, a fast convergence
to a fixed point, such that $x=f(x)$, is observed
(left panel of Fig.~\ref{fig-fx5-10}). Raising $\lambda$,
a different behaviour is observed: for $\lambda=10$, the process
converges to a limit cycle, corresponding to the oscillation between
two distinct values of $x$ (right panel of Fig.~\ref{fig-fx5-10}).
Further increasing $\lambda$, a second qualitative change of behaviour
appears, as seen for $\lambda=15$. Here, no regular pattern is observed,
and the process becomes chaotic, as seen on the left panel of
Fig.~\ref{fig-fx15}.
Extending the time window (right panel), one sees that the irregular
behaviour is indeed a stationary feature, and not simply a transient effect.

Note that this transition from fixed points to chaotic behaviour
through limit cycles when varying the control parameter
is believed to be a generic scenario for the transition to chaos
\cite{Berge}.

\subsection{Chaotic walk}

As discussed in section~\ref{sec-stoch}, a paradigmatic stochastic model
is the random walk, which has many applications in different fields.
It is interesting to observe that the above chaotic map can be used
to define a simple deterministic analog of the random walk,
namely a chaotic walk \cite{chwalk}.

\begin{figure}
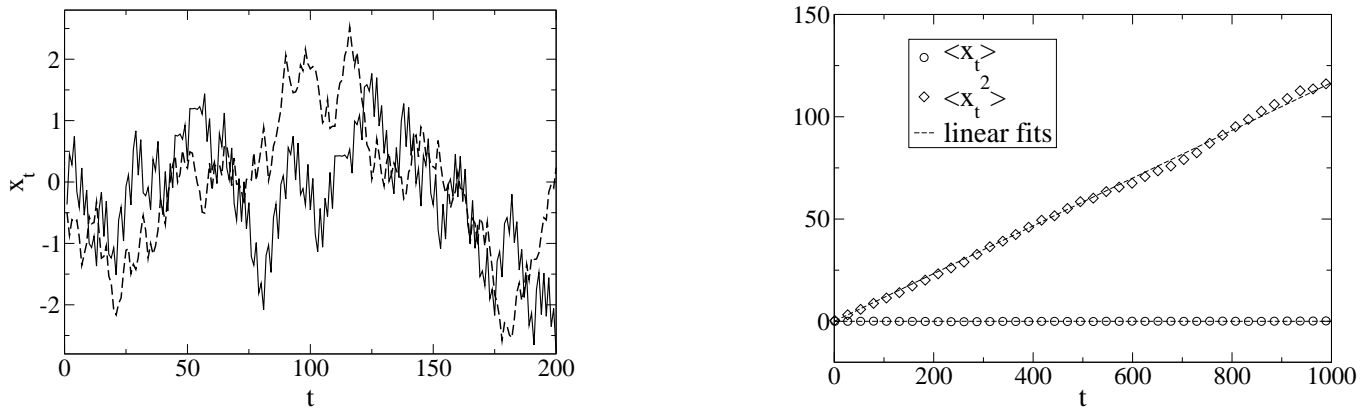

\includegraphics[width=0.42\textwidth,clip]{chrw.eps}
\hfill
\includegraphics[width=0.42\textwidth,clip]{chx2.eps}
\caption{Left: Chaotic walk (full line); a random walk (dashed line)
is shown for comparison.
Right: Mean displacement $\la x_t \ra$ and mean square displacement
$\la x_t^2 \ra$ of the chaotic walk, obtained by averaging over many
trajectories having different initial conditions. The obtained results
are very similar to the results that would be obtained for a random walk.}
\label{fig-chw}
\end{figure}

Let us start by introducing a variable $u_t$ evolving according to
\be
u_{t+1}=f(u_t), \qquad 0 \le u_t \le 1,
\ee
where $f(u)$ is the function defined in Eq.~(\ref{eq-map-fx}).
The idea is now to use the chaotic variable $u_t$ as the increment
of a walk, namely
\be
x_{t+1} = x_t + (2u_t-1).
\ee
Note that, more precisely, the increment $(2u_t-1)$ is used instead of $u_t$
in order for the increment to take values between $-1$ and $1$,
that is, on a symmetric interval around $0$.
The chaotic walk is illustrated in the left panel of Fig.~\ref{fig-chw}.
The visual similarity with a random walk, shown for comparison,
is clear. A closer look however reveals
a short time anticorrelation in the chaotic walk: positive
increments are more likely to be followed by negative increments than
by positive ones.

A more quantitative comparison is obtained by computing the mean
displacement $\la x_t \ra$ and the mean square displacement $\la x_t^2 \ra$.
Average values are obtained as ensemble averages, that is by averaging over a large
set of trajectories obtained by varying the initial conditions
(averages are thus time-dependent).
Here, the initial position is fixed to $x_0=0$, and the initial increment
$u_0$ is sampled in a uniform way from the interval $(-1,1)$.
Note that the sampling is deterministic, that is, equidistant values
on the interval $(-1,1)$ are chosen.
The resulting mean displacement $\la x_t \ra$ is found to be almost equal
to zero, up to the measurement uncertainty, as seen on
the right panel of Fig.~\ref{fig-chw}.
The mean square displacement $\la x_t^2 \ra$ is found to be linear
in time, as would be the case for a random walk.
Hence, one sees that such simple and standard indicators as mean
and mean square displacements cannot be used to discriminate between
chaotic and deterministic signals. In addition, it also suggests that
from a practical perspective, the distinction between the deterministic
or stochastic character of a signal might not be so important.
Indeed, if a deterministic system generates a signal similar to the
above chaotic walk, why not modeling it as a random walk?
This very question is the topic of the next section.

\section{Stochasticity vs.~determinism: a choice of description?}
\label{sec-compar}

\begin{figure}
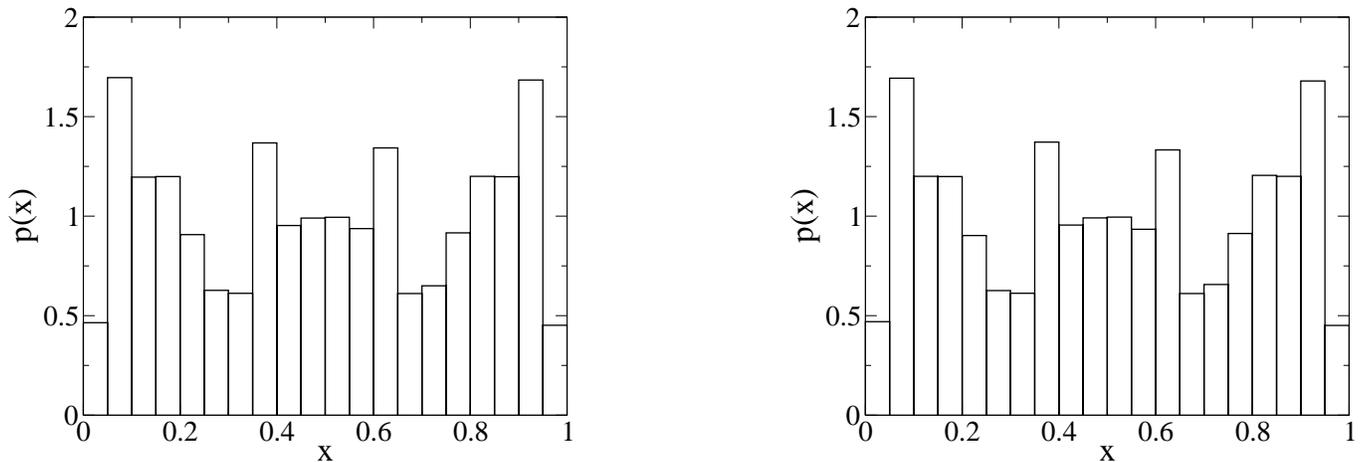

\includegraphics[width=0.42\textwidth,clip]{chhisto.eps}
\hfill
\includegraphics[width=0.42\textwidth,clip]{ranhisto.eps}
\caption{Left: Histogram of the values of $x_t$, using the deterministic
evolution $x_{t+1}=f(x_t)$, in the case $\lambda=15$.
Right: Histogram obtained from the effective stochastic process
mimicking the deterministic one (see text).
Both histograms are hardly distinguishable.}
\label{fig-ch-histo}
\end{figure}

Let us come back to the map $x_{t+1}=f(x_t)$ studied in
section~\ref{sec-map-fx}. Instead of simply computing a few average
values, a richer information is obtained by determining the histogram
of the values of $x_t$.
The resulting histogram is shown on the left panel of Fig.~\ref{fig-ch-histo},
in the case $\lambda=15$ where the dynamics is chaotic.

As usual, the histogram is build by splitting the interval $(0,1)$
into a certain number of bins, and by counting how many values
of $x_t$ fall into a given bin, along the trajectory.
Now it is interesting to note that the dynamics is deterministic
on condition that the value of $x_t$ is known with an infinite precision.
However, in practice, only a finite precision on the data is available.
To illustrate this issue, let us consider that we use the above bins
not only to determine the histogram, but also to simulate the dynamics
of $x_t$, that is, to define the ``microscopic'' configurations of
the process. Then, in terms of bins, the dynamics can no longer be
considered as purely deterministic, as the value of $x_t$ in a given
bin can evolve into different bins as time elapses. In other words,
the choice of an initial bin is not enough to determine the occupied bin
at any later time.

An effective stochastic dynamics can then be defined in the following way.
First, we measure the transition probability $T(j \to k)$ from any
bin $j$ to any bin $k$ in a single time step ('probability' is here
understood as a frequency of occurrence --see the paper by J.~Velasco
in this Special Issue).
Then, in a second step, we simulate a stochastic process defined by the
transition probabilities $T(j \to k)$ measured in the coarse-grained
deterministic process.
The histogram of this stochastic process can also be measured,
and can be compared with the original histogram of the deterministic
dynamics (see Fig.~\ref{fig-ch-histo}). A striking similarity is observed,
showing again that in practical situations, it may be hard to distinguish
a stochastic process from a deterministic one. Though this result might be
understood as a negative statement, such a property also has some
advantages, as it might be convenient in some cases to model
a deterministic process by a stochastic one.

There are however more sophisticated ways to try to distinguish between
a deterministic (chaotic) and random signals.
One can look for instance for the dimension of the underlying attractor
(the generalization of the notion of fixed points, having zero dimension,
or of limit cycle, having dimension one), assuming implicitly the process
to be deterministic. The dimension found for a stochastic process
would then in principle be infinite.
However, as for the other (more naive) indicators discussed above,
the stochastic or deterministic nature of a process may
remain very difficult to assess on the basis of real data,
which basically consist in a finite set of points.
With this issue in mind, some authors have formulated the interesting
proposition to classify the behaviour of a signal as ``stochastic
or deterministic on a certain scale of resolution'' \cite{Cencini}.

\section{Conclusion}

As a summary, we have tried to give in these notes a brief pedagogical
account of some similarities and differences between stochastic
and deterministic processes, providing along the way a short presentation
of some basic notions on the mathematical description of stochastic
processes. Our aim was to illustrate that the gap between the behaviour
of stochastic and deterministic processes might, quite surprisingly,
be much smaller than a priori expected.

Let us emphasize that there indeed exists an absolute distinction,
at the conceptual level, between deterministic and stochastic processes.
These two types of processes obey different types of laws or equations,
and can thus be distinguished without ambiguity when dealing with the
mathematical formalism.
However, from a practical perspective, that is when dealing with real data
(either coming from a real-world experiment, or from numerical simulations),
one has to cope with the finite resolution of the data, and the distinction
between stochasticity and determinism becomes blurred
--at least on the basis of the sole data.
Of course, it may also be known that the data were obtained
through a deterministic numerical simulation, but this knowledge constitutes
an extra piece of information, not contained in the data themselves.
Hence, to some extent, determinism or stochasticity are, at a practical level,
choices of description (``do I use a deterministic or a stochastic model
to describe a given real system?'')

As a final remark, let us also note that even at a conceptual level,
determinism and stochasticity are notions that apply only to
\emph{mathematical descriptions}, that is, to models of the
real world, and not to the real world itself. It is thus not clear if
asking whether some data obtained from a real-world experiment
are intrinsically deterministic or stochastic is a meaningful question.
One should probably rather ask which type of model, deterministic
or stochastic, is the more relevant to describe the data.

\end{document}